\documentclass[preprint,aps,prb,nofootinbib,floatfix,superscriptaddress]{revtex4-2}
\usepackage[utf8]{inputenc}

\usepackage{amsmath}    
\usepackage{graphicx}   
\usepackage{verbatim}   
\usepackage{color}      
\usepackage{epstopdf}  
\usepackage{upgreek} 
\usepackage[hidelinks]{hyperref}   
\usepackage{textcomp} 
\usepackage[separate-uncertainty=true, multi-part-units=single]{siunitx}
\usepackage{dcolumn}
\usepackage[capitalise]{cleveref}

\newcommand{\Ga}{Ga$^{+}$}
\newcommand{\He}{He$^{+}$}
\newcommand{\ie}{\textit{i.e.}}

\newcommand{\ea}{\textit{et al.}}
\DeclareMathOperator\erf{erf}

\begin{document}

\title{Controlling magnetic skyrmion nucleation with Ga$^{+}$ ion irradiation}

\author{Mark C.H. de Jong}
\email{m.c.h.d.jong@tue.nl}
\affiliation{Department of Applied Physics, Eindhoven University of Technology, P.O. Box 513, 5600 MB Eindhoven, the Netherlands}
\author{Bennert H.M. Smit}
\affiliation{Department of Applied Physics, Eindhoven University of Technology, P.O. Box 513, 5600 MB Eindhoven, the Netherlands}
\author{Mari\"{e}lle J. Meijer}
\affiliation{Department of Applied Physics, Eindhoven University of Technology, P.O. Box 513, 5600 MB Eindhoven, the Netherlands}
\author{Juriaan Lucassen}
\affiliation{Department of Applied Physics, Eindhoven University of Technology, P.O. Box 513, 5600 MB Eindhoven, the Netherlands}
\author{Henk J.M. Swagten}
\affiliation{Department of Applied Physics, Eindhoven University of Technology, P.O. Box 513, 5600 MB Eindhoven, the Netherlands}
\author{Bert Koopmans}
\affiliation{Department of Applied Physics, Eindhoven University of Technology, P.O. Box 513, 5600 MB Eindhoven, the Netherlands}
\author{Reinoud Lavrijsen}
\email{r.lavrijsen@tue.nl}
\affiliation{Department of Applied Physics, Eindhoven University of Technology, P.O. Box 513, 5600 MB Eindhoven, the Netherlands}

\date{\today}

\begin{abstract}
In this paper, we show that magnetic skyrmion nucleation can be controlled using {\Ga} ion irradiation, which manipulates the magnetic interface effects (in particular the magnetic anisotropy and Dzyaloshinskii-Moriya interaction) that govern the stability and energy cost of skyrmions in thin film systems. We systematically and quantitatively investigated what effect these changes have on the nucleation of magnetic skyrmions. Our results indicate that the energy cost of skyrmion nucleation can be reduced up to 26${\%}$ in the studied dose range and that it scales approximately linearly with the square root of the domain-wall energy density. Moreover, the total number of nucleated skyrmions in irradiated devices after nucleation was found to depend linearly on the ion dose and could be doubled compared to non-irradiated devices. These results show that ion irradiation cannot only be used to enable local nucleation of skyrmions, but that it also allows for fine control of the threshold and efficiency of the nucleation process.
\end{abstract}

\maketitle
Magnetic skyrmions are a type of chiral magnetic texture that has generated tremendous research interest in recent years \cite{Wiesendanger2016, Fert2017, Everschor-Sitte2018, Zhang2020, Tokura2021}. Due to their small size, down to a few nanometers \cite{Tokura2021}, current induced motion \cite{Woo2016}, stability and topology \cite{Je2020} they are promising candidates for future memory applications \cite{Tomasello2014, Luo2021} as well as computing schemes such as neuromorphic computing \cite{Grollier2020}. Stable skyrmions at room temperature can be readily stabilized and observed in magnetic multilayers, as was shown in Refs. \cite{Woo2016, Boulle2016, Moreau2016}. A skyrmion in such a system is present in each of the magnetic layers, forming a tube in three dimensions. In a multilayer, skyrmions are stabilized primarily by dipolar interactions which are enhanced due to the large magnetic volume of all the layers combined \cite{Buettner2018}. A uniform chirality in each layer---which distinguishes skyrmions from magnetic bubbles---is realized by a strong interfacial Dzyaloshinskii-Moriya interaction (DMI) \cite{Dzyaloshinskii1958, Moriya1960} which is present and strong in these systems due to the large number of complementary heavy metal $|$ magnet interfaces \cite{Moreau2016}. Skyrmions can be nucleated in such systems in a variety of ways: using magnetic fields that break larger domains down to skyrmions \cite{Soumyanarayanan2017, Fallon2020, Juge2021}, spin-orbit torques at defects \cite{Buettner2017, Kern2022}, and thermal fluctuations of the magnetization caused by either ultrafast laser pulses \cite{Je2018, Buettner2021} or Joule heating during nanosecond current pulses \cite{Legrand2017, Lemesh2018}.  

In all the above-mentioned cases, the nucleation of skyrmions occurs at random positions in the material. For many applications however, it is desirable to be able to nucleate skyrmions at well-defined positions in a device and to control the amount of excitation that is needed to nucleate a skyrmion. Many schemes to achieve this have been experimentally demonstrated or proposed \cite{Zhang2020, Finizio2019, Ohara2021, Everschor-Sitte2017, Buettner2017, Fallon2020, Juge2021, Kern2022} but in this paper we focus on the use of {\Ga} ion irradiation. 

Ion irradiation can be used to locally tune the magnetic effects originating from interfaces by gradually increasing the degree of intermixing \cite{Devolder2000, Hyndman2001, Vieu2002} in systems consisting of single magnetic layers \cite{Devolder2000, Hyndman2001, Vieu2002, Juge2021} and even magnetic multilayers \cite{Fallon2020, DeJong2022, Kern2022, Hu2022}. In the case of field-driven skyrmion nucleation, localized \cite{Fallon2020, Hu2022} as well as non-localized \cite{Hu2022} {\Ga} ion irradiation has been shown to facilitate field driven skyrmion nucleation. Both studies found that skyrmions nucleate more densely in regions that have been irradiated with {\Ga} ions and were able to nucleate skyrmions at well defined positions with an applied magnetic field. Recently, there has also been work studying the effect of localized {\He} ion irradiation on the field driven \cite{Juge2021}, as well as current- and laser-induced nucleation of skyrmions \cite{Kern2022} and also their current driven motion \cite{Juge2021, Kern2022}. Skyrmion nucleation was achieved exclusively within the irradiated regions of the devices, demonstrating a reduced threshold for nucleation compared to nonirradiated regions. It was also shown that the border between the irradiated and nonirradiated regions presents an energy barrier for skyrmions. However, the cause of the decrease in the nucleation threshold as well as the dependence of the nucleation threshold on the ion dose are still unclear and the extent to which the nucleation threshold can be reduced remains an open question.


We have previously studied the effect that {\Ga} ion irradiation has on the magnetic properties of an [Ir(1)$|$Co(0.8)$|$Pt(1)]$_{\times 6}$ multilayer system \cite{DeJong2022} (all thicknesses in brackets are in \SI{}{nm}) and found that the effective anisotropy $K_{\text{eff}}$ decreases gradually and strongly throughout the tested dose range and that the interfacial DMI strength $D$ decreases gradually but much less strongly. The effect of these changes is a factor 2 reduction of the domain-wall energy density,
\begin{equation}
\sigma_{\text{DW}} = 4 \sqrt{A K_{\text{eff}}} - \pi |D|,
\end{equation}
\noindent where $A$ is the exchange stiffness. Despite this large decrease, the relatively small decrease in the strength of the DMI ensures that domain walls remain chiral. Hence, we postulated that {\Ga} ion irradiation would be a convenient tool to manipulate the energy barrier for skyrmion nucleation.


In this paper, we study current-driven nucleation of magnetic skyrmions in the exact same material system as was used in Ref. \cite{DeJong2022} at different {\Ga} ion doses. By irradiating the entire device we could study the effect of ion irradiation on the current-driven nucleation of skyrmions in a statistical manner and determine the dependence of the threshold current for skyrmion nucleation and skyrmion density as a function of ion dose. We found that ion irradiation decreases the threshold current for nucleation and that the dependence on ion dose of the threshold current resembles the dependence of the domain-wall energy density on the ion dose, showing that the change in the magnetic parameters is an important contribution to the modification of the threshold current. Surprisingly, we also found that the skyrmion density in our devices scales linearly with ion dose, doubling in the tested dose range, which enables control of the skyrmion density in devices by ion irradiation.

\section{Methods}
The material stack used throughout this paper is the following magnetic multilayer: Si$|$SiO$_{2}$(100)$||$Ta(4)$|$Pt(15)$|$[Ir(1)$|$Co(0.8)$|$Pt(1)]$_{\times 6}$$|$Pt(2). All layers were grown using DC magnetron sputtering in an argon atmosphere with partial pressure of \SI{2e-3}{mbar}. The system had a base pressure of $\sim$\SI{5e-9}{mbar}. The nucleation devices were patterned using electron beam lithography in combination with lift-off and consist of a \SI{7}{\micro\meter}long and \SI{1.5}{\micro\meter} wide strip [see \cref{fig:Figure_1}(a)]. {\Ga} ion irradiation of this strip was performed using the {\Ga} focussed ion beam in a FEI Nova NanoLab Dualbeam, with an acceleration voltage of \SI{30}{keV} and a beamcurrent of approximately \SI{1.5}{pA}. To irradiate areas larger than the beam diameter the beam was rasterscanned across the sample. The dose $d$ was calculated as:
\begin{equation}
d = \frac{BC \times DT}{LS \times SS},
\end{equation} 
where $BC$ is the beamcurrent, $DT$ is the dwell time at each position and $LS$ and $SS$ are the line and spot spacing, respectively. The line and spot spacing were limited by the minimum dwell time and maximum beam speed and were set to \SI{80}{nm} for all devices used in this paper. It is likely that the spot size of the ion beam was smaller than the spot and line spacing during irradiation since at optimal focus the system has a resolution of approximately \SI{20}{nm} \cite{Franken2011}. Nevertheless, the spot and line spacing is smaller than all skyrmions that were observed and no obvious square pattern is present in the nucleation sites (see Supplemental Material S4 \cite{NoteX}; see, also, Refs. \cite{Scarioni2021,Tan2020,Buettner2018} therein]).

The fabricated and irradiated devices were mounted in a custom sample holder positioned in a Br{\"u}ker Dimension Edge atomic force microscope that was used for magnetic force microscopy (MFM) measurements using custom-coated low moment tips (Ta(4) $|$ Co(7.5) $|$ Ta(5) sputter deposited onto Nanosensors PPP-FMR tips). The sample holder enables manipulation of the magnetic state of the multilayers using nanosecond current pulses, as well as the simultaneous application of a magnetic field up to $\mu_{\text{0}}H_{\text{z}} =$ \SI{500}{mT} during measurements. Skyrmion nucleation was achieved by sending bipolar pulse trains of nanosecond current pulses using an Agilent 33250A \SI{80}{MHz} arbitrary waveform generator. The pulses were measured using an Agilent Infiniium DSO80604B oscilloscope connected in series with the nucleation devices. The current through the device was calculated by dividing the pulse voltage as measured with the oscilloscope by the resistance of the oscilloscope (\SI{50}{\ohm}). Finally, the current density was calculated by assuming that the current is uniformly distributed throughout the entire multilayer stack.

\section{Results and Discussion}
\begin{figure*}
\includegraphics[width=\textwidth]{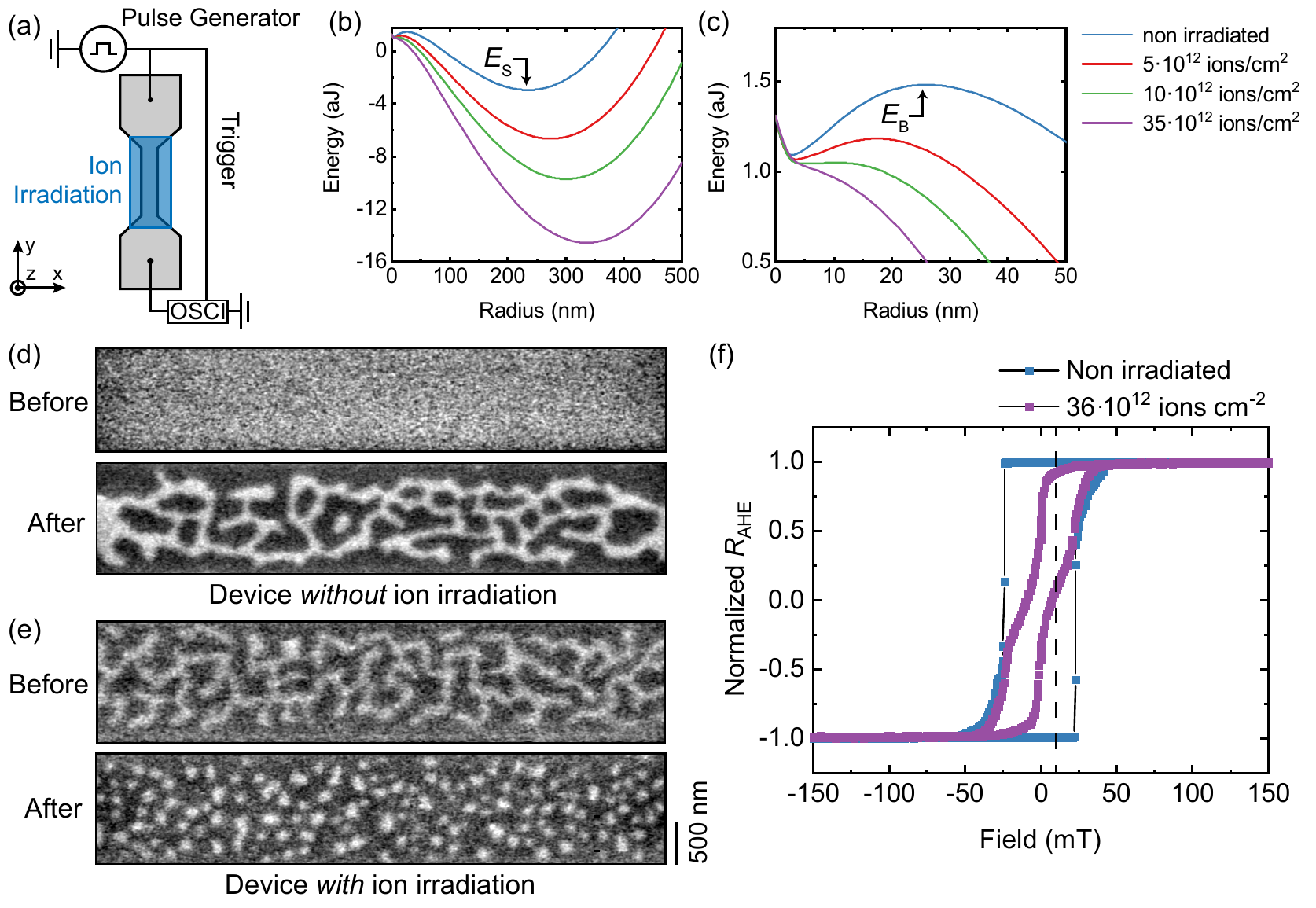}
\caption{\label{fig:Figure_1}(a) Schematic overview of the nucleation devices (not to scale). Each device has been irradiated with a different {\Ga} dose in the region indicated in blue. (b) Energy as a function of skyrmion radius calculated using the model by B{\"u}ttner {\ea} from Ref. \cite{Buettner2018} for the magnetic parameters corresponding to different doses of {\Ga} ion irradiation. (c) Zoom of the energy barrier for skyrmion nucleation. (d) MFM images of the domain state in a non-irradiated device before and after applying a current pulse train with pulses of $J=$ \SI{5.54e11}{A.m^{2}} in a magnetic field of $\mu_{0}H_{\text{z}}=$ \SI{10}{mT}. (e) MFM images of the magnetic state in a device irradiated with a dose of $d=35\times10^{12}$ ions \SI{}{cm^{-2}}. The current density, pulse shape, number of pulses and magnetic field are all the same as in (d). (f) Hysteresis loops measured in a similar non irradiated device and a device irradiated with $d=36\times10^{12}$ ions \SI{}{cm^{-2}} using the anomalous Hall effect. The dashed line corresponds to the magnetic field applied in (d) and (e). These devices have a thinner Pt(\SI{2}{nm}) buffer layer but are otherwise nominally identical.}
\end{figure*}

To motivate why the nucleation of skyrmions should be affected by a decrease in the domain-wall energy density we consider a model of the skyrmion energy from Ref. \cite{Buettner2018}. In \cref{fig:Figure_1}(b) and (c) we plot the energy of a skyrmion as a function of its radius for the structural and magnetic parameters measured for our (irradiated) Ir$|$Co$|$Pt multilayer system in Ref. \cite{DeJong2022}. For increasing ion dose we indeed observe that the local energy maximum $E_{\text{B}}$ [\cref{fig:Figure_1}(c)] that prevents nucleation of a skyrmion, and is related to the competition between the domain-wall energy (and the Zeeman energy) with the dipolar energy, decreases in magnitude. Although this model of a nucleation process is likely oversimplified, we still expect a decrease in the domain-wall energy density to affect the nucleation since this always includes an increase in the domain-wall length. The model also predicts a decrease in the energy minimum $E_{\text{S}}$ [\cref{fig:Figure_1}(b)] which means that the energy of a skyrmion is lower in an irradiated part of the sample than a nonirradiated part. This matches the observation in Ref. \cite{Juge2021} that the border between these regions represents an energy barrier for a skyrmion. These results suggest that {\Ga} ion irradiation should indeed affect the nucleation of magnetic skyrmions.

In order to experimentally study the dependence of the current-driven skyrmion nucleation on the {\Ga} dose, we utilized devices shown schematically in \cref{fig:Figure_1}(a). First, the magnetization in the device was saturated by applying a magnetic field of $\mu_{0}H_{\text{z}} =$\SI{-150}{mT}. Next, a small bias field $\mu_{0}H_{\text{z}} = $\SI{10}{mT} with the opposite polarity was applied for the duration of the measurement. To nucleate skyrmions, we applied a pulse train consisting of 1000 bipolar nanosecond current pulses with a current density in the order of $10^{11}$ \SI{}{A/m^{2}} (pulse length: \SI{50}{ns}, time between pulses: \SI{1}{\micro\second}). This is a similar nucleation scheme as used in Refs. \cite{Legrand2017, Lemesh2018}, where the nucleation is primarily driven by Joule heating. This increase in temperature drives fluctuations in the magnetization that result in the formation of topological charge and hence skyrmions \cite{Lemesh2018}. We used this method of nucleation because the relative change in Joule heating required for skyrmion nucleation can be easily determined by measuring the threshold current density for nucleation and the resistance of the devices as a function of ion dose.


Two examples that show that {\Ga} ion irradiation indeed had a large effect on the nucleation of skyrmions are shown in \cref{fig:Figure_1}(d) and (e). In \cref{fig:Figure_1}(d), we show MFM images of the magnetic domain state in a non-irradiated device before and after applying a pulse train with a current density of $J=$ \SI{5.54e11}{A.m^{2}} (bias field parallel to the dark domain). Only one skyrmion was present in the device after applying the pulses. However, when we applied the same pulse train and magnetic field to a device that had been irradiated with $d=35\times10^{12}$ ions \SI{}{cm^{-2}}, we found that the device was in a state where only skyrmions remained [\cref{fig:Figure_1}(e)]. We also show two anomalous Hall effect measurements that were done as a function of magnetic field in two similar devices in \cref{fig:Figure_1}(f) to illustrate the effect of the ion irradiation on the magnetic hystresis loops. The loop changed from a square loop in the nonirradiated device to a slanted loop in the irradiated device. In the remainder of this paper we study the effect on nucleation in more detail and examine how strongly the nucleation of skyrmions can be controlled.

\begin{figure}
\includegraphics{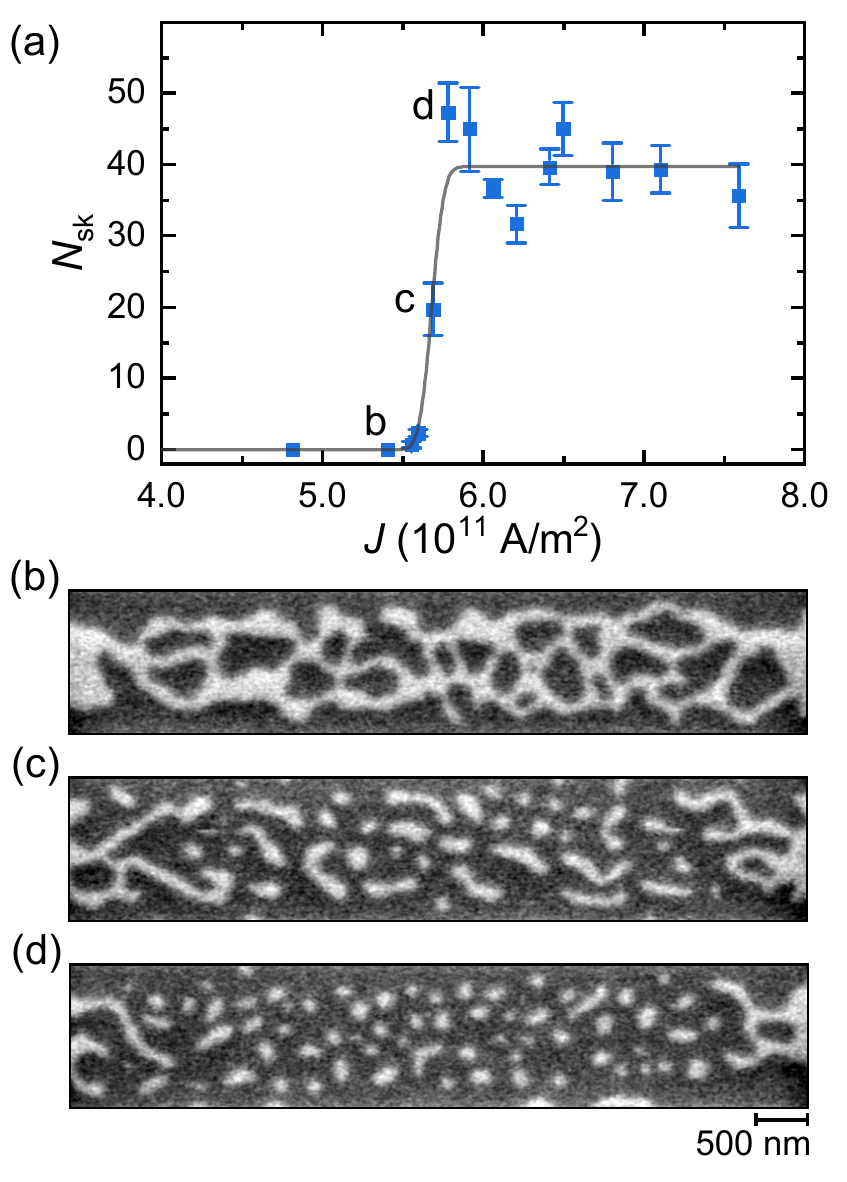}
\caption{\label{fig:Figure_3}(a) The number of skyrmions that appear in a device after applying a sequence of 1000 bipolar current pulses with a length of \SI{50}{ns} and a time between pulses of \SI{1}{\micro\second} plotted as a function of the applied current density. Each measurement point is an average of 3 nucleation events, error bars represent one standard deviation. The gray line is a fit to the date with the error function, \cref{eq:error_function}. (b-d) MFM images of the magnetization state in the device after applying the pulses, the corresponding current densities have been labelled in (a). Dark contrast is parallel to the applied bias field.}
\end{figure}


To do this we require a measure for the energy required to nucleate skyrmions in our devices. The threshold current density is a logical choice but the transition between states with no skyrmions and only skyrmions is not infinitely sharp and therefore it is not enough to only look at whether one or more skyrmions is present in the device. Hence, we measured the number of skyrmions in the device after applying a train of current pulses as a function of current density. All measurements were performed with the same bias field of $\mu_{0}H_{\text{z}}=$ \SI{10}{mT} (in the Supplemental Material SII we show that skyrmions were nucleated in both nonirradiated devices and irradiated devices for this bias field.) To obtain the amount of nucleated skyrmions we used a MatLab script which is based on a skyrmion counting procedure described in Ref. \cite{Tan2020} and is described in the Supplemental Material SIII. For each current density three nucleation events were observed and the number of skyrmions was averaged. As a measure for the uncertainty in the final number of skyrmions we took the standard deviation of this average.


The results for a nonirradiated device are shown in \cref{fig:Figure_3}(a) where the number of skyrmions is plotted as a function of current density. For low current densities ($J<$ \SI{5e11}{A.m^{-2}}) the effect on the magnetization was not strong enough to nucleate skyrmions and no skyrmions appear in the device [\cref{fig:Figure_3}(b)]. For high current densities ($J>$ \SI{6e11}{A.m^{-2}}) on average 40 skyrmions were nucleated in the device [\cref{fig:Figure_3}(d)]. In between, a sharp transition was found where a threshold current for nucleation is exceeded. To determine this threshold current accurately we fit the data with the following phenomenological model:
\begin{equation}\label{eq:error_function}
N_{\text{sk}} = \frac{1}{2} N_{\text{sk,sat}} \Big( 1 + \erf \Big( \frac{J - J_{\text{c}}}{\sqrt{2} \, \Delta} \Big) \Big),
\end{equation}
\noindent where $J_{\text{c}}$ is the threshold current, $N_{\text{sk}}$ ($N_{\text{sk,sat}}$) is the number of skyrmions (above the threshold current), $\Delta$ is the width of the transition region and erf() is the error function. The threshold current is found to be $J_{\text{c}} = (5.68 \pm 0.02) \times 10^{11}$ \SI{}{A.m^{-2}} for the nonirradiated device, where we have taken the fit uncertainty as the error in the threshold current. Additionally, we can now conclude that the width of the transition region $\Delta = (0.06 \pm 0.02) \times 10^{11}$ \SI{}{A.m^{-2}} is much less than the threshold current, {\ie} the transition is sharp. Now that we have established a method to reliably determine the threshold current needed for the nucleation of skyrmions we can look at the effect of ion irradiation on the threshold current.

\begin{figure*}[p!]
\includegraphics[width=\textwidth]{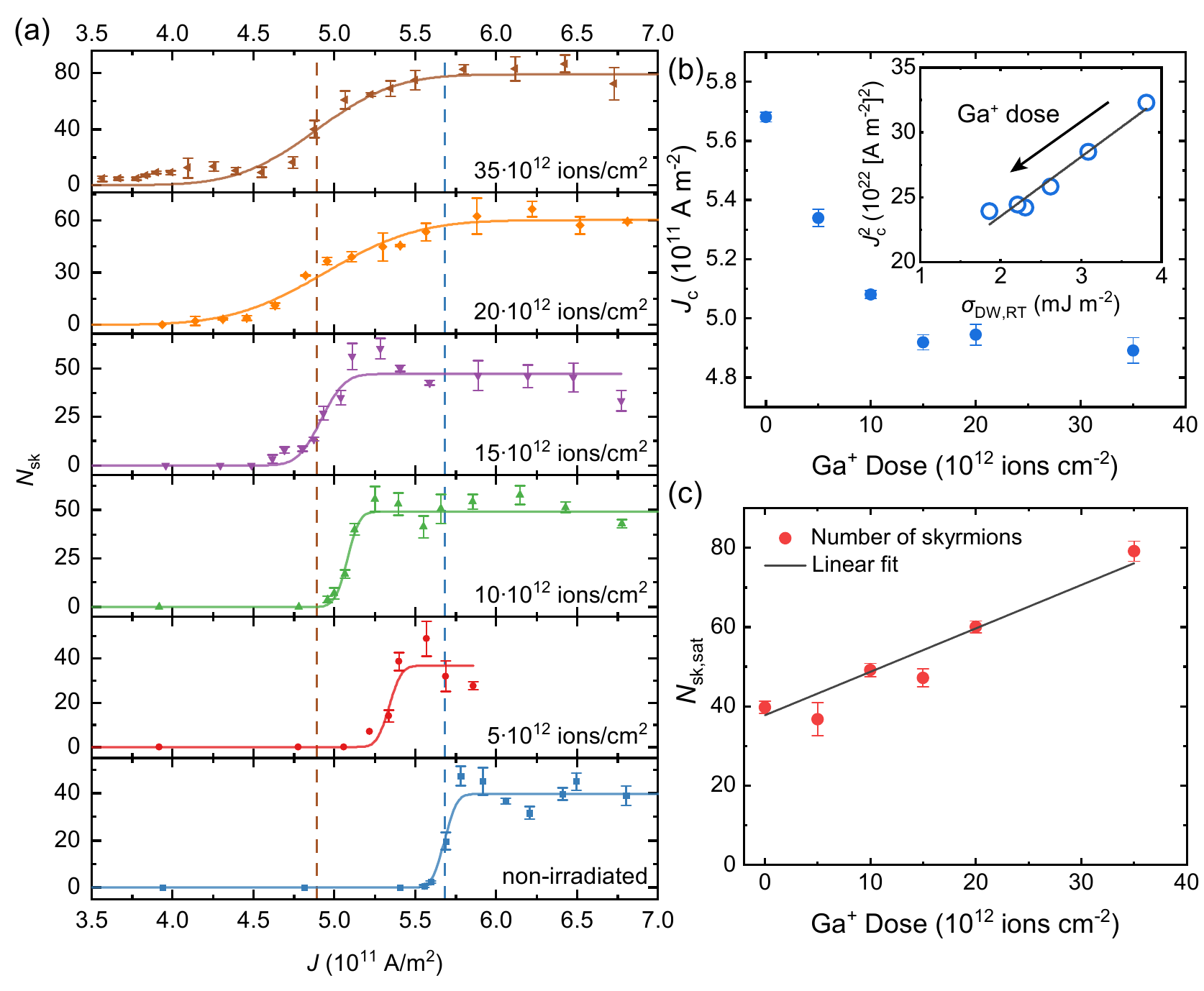}
\caption{\label{fig:Figure_4}(a) The number of skyrmions in a device after applying a sequence of 1000 bipolar current pulses plotted as a function of the applied current density for several devices irradiated with different {\Ga} doses. The lines are fits to the data with the error function, \cref{eq:error_function}. (b) The threshold current $J_{\text{c}}$, extracted from the data in (a), plotted as a function of the {\Ga} dose. Inset: The threshold current squared plotted as a function of the domain-wall energy density $\sigma_{\text{DW,RT}}$, measured for the same material system at room temperature in Ref. \cite{DeJong2022}. The dark grey line is a linear fit to the data. (c) Average number of skyrmions that appear in the device for current densities larger than the threshold current plotted as a function of {\Ga} dose. The dark grey line is a linear fit to the data.}
\end{figure*}

The experiment highlighted in \cref{fig:Figure_3}(a) was repeated for several ion doses up to $d=35\times10^{12}$ {\Ga} ions \SI{}{cm^{-2}}. The results are plotted in \cref{fig:Figure_4}(a) for six different ion doses between zero and $d = 35 \times 10^{12}$ ions \SI{}{cm^{-2}}. Each dataset is fitted with \cref{eq:error_function}, which is in good agreement with the data for all doses. We begin by examining the change in the threshold current $J_{\text{c}}$ as a function of ion dose. In \cref{fig:Figure_4}(a) the two dotted lines represent the threshold current for the nonirradiated device (blue squares) and the device with the highest ion dose (brown triangles pointing left). It is apparent that the ion irradiation has significantly lowered the threshold current required for skyrmion nucleation. To determine the dependence in more detail we have plotted the threshold current against the {\Ga} ion dose in \cref{fig:Figure_4}(b). For doses up to $d = 15 \times 10^{12}$ ions \SI{}{cm^{-2}} the threshold current reduces monotonously as a function of dose. At this dose the decrease stops and for larger doses the threshold current appears to saturate at a value of $J_{\text{c}} = (4.92 \pm 0.06) \times 10^{11}$ \SI{}{A.m^{-2}}. This trend is similar to the trend we observed for the domain-wall energy density as a function of ion dose \cite{DeJong2022}. In the Supplemental Material SI we also show that the observed change in the threshold current is much larger than the naturally occurring variation in threshold current between devices indicating that the observed reduction in threshold current is a result of the ion irradiation.


We assume that a thermal mechanism (as in Refs. \cite{Legrand2017, Lemesh2018}) is also the dominant mechanism for nucleation in the system used here. Evidence for this is the observation that the required current density for nucleation decreased when the pulse length was increased (compare Fig. S2(a) in the Supplemental Material with \SI{10}{ns} pulses to \cref{fig:Figure_3}(a) with 50 ns pulses.). This is expected for thermally-driven nucleation but not for SOT-driven nucleation, since this should happen at timescales on the order of \SI{1}{ns} \cite{Lemesh2018}. With this assumption, the energy that is deposited into the system scales with $J^{2}$. Hence, the relative difference in energy requirement for the nonirradiated device and the device irradiated with $d = 35 \times 10^{12}$ ions \SI{}{cm^{-2}} can be calculated as $(J_{\text{c, d=0}}^{2} - J_{\text{c, d=35}}^{2}) / J_{\text{c, d=0}}^{2} = 0.26$, {\ie} the energy required for nucleating skyrmions is reduced by $26{\%}$ \footnote{The resistance increases slightly with ion dose as well, but is only 4 ${\%}$ higher in the device irradiated with $d = 35 \times 10^{12}$ ions \SI{}{cm^{-2}} compared to the nonirradiated device and hence this is not enough to explain the decrease in the required Joule heating for nucleation.}.

In the inset of \cref{fig:Figure_4}(b) we have also plotted the measured threshold current squared as a function of the measured domain-wall energy density at room temperature from Ref. \cite{DeJong2022}. We find that for doses up to $d = 15 \times 10^{12}$ ions \SI{}{cm^{-2}} there is an approximately linear dependence of $J_{\text{c}}^{2}$ on $\sigma_{\text{DW,RT}}$ as expected when Joule heating is the dominant mechanism for nucleation. However, the range of domain-wall energies that we were able to probe with ion irradiation was too small to conclusively determine the relation between the critical current and the domain-wall energy density.

A second observation, is that the number of skyrmions that appeared in the bar for current densities larger than the threshold current increased with increasing ion dose. This behavior is clear when we plot the number of skyrmions at saturation $N_{\text{sk,sat}}$ as a function of the ion dose in \cref{fig:Figure_4}(c). In the nonirradiated device there was an average of 40 skyrmions in the device after applying a sufficiently strong current pulse train. This number increased linearly with ion dose to 80 skyrmions for the device with $d = 35 \times 10^{12}$ ions \SI{}{cm^{-2}}, a doubling of the total number of skyrmions. An increase of the skyrmion density with increasing {\Ga} ion dose has also been observed in field-driven nucleation experiments \cite{Hu2022} and is expected when reducing the domain-wall energy density \cite{Soumyanarayanan2017}. Considering that the decrease in the domain-wall energy density as a function of dose in our samples slows down for doses larger than $d = 15 \times 10^{12}$ ions \SI{}{cm^{-2}} \cite{DeJong2022} a linear increase in the number of skyrmions cannot be explained by a decrease in the domain-wall energy alone, which could suggests that the number of nucleation sites in the device also increases as a function of {\Ga} ion dose.


One final observation we can make from \cref{fig:Figure_4}(a) is that for higher ion doses the width of the transition region between no skyrmions and saturation increased significantly. The good agreement between the data and \cref{eq:error_function} for all doses indicates that the nucleation current followed a normal distribution with $\Delta$ the standard deviation, which broadens significantly for doses above $d = 15 \times 10^{12}$ ions \SI{}{cm^{-2}}. At this dose, we also observed that the domain-wall energy stops decreasing as a function of ion dose and saturated \cite{DeJong2022}. This suggests that at higher doses the variance in the magnetic parameters might be more strongly affected by {\Ga} ion irradiation than the average value, {\ie} the local variation of magnetic interface effects in the material increases. This hypothesis is also in line with an increase in the number of nucleation sites as postulated in the previous paragraph. However, verifying this hypothesis by measuring the magnitude of such local variations in the magnetic interface parameters as a function of ion dose is beyond the scope of this paper.


\section{Conclusion and Outlook}
In this paper, we have reported the effect of {\Ga} ion irradiation on the current-driven nucleation of skyrmions in a magnetic multilayer. We uniformly irradiated our devices which enabled us to study the effect that changes in the magnetic parameters due to the irradiation have on the current-driven nucleation process, contrary to earlier work where the authors primarily looked at nucleation from small irradiated regions \cite{Fallon2020, Juge2021, Kern2022}. We found that the threshold current for skyrmion nucleation can be reduced for {\Ga} doses up to $d = 15 \times 10^{12}$ ions \SI{}{cm^{-2}} and that for larger doses the threshold current saturates. Additionally, and perhaps more importantly for applications where precise control of the skyrmion nucleation sites is desired, for doses up to $d = 15 \times 10^{12}$ ions \SI{}{cm^{-2}} there exists a current density where the number of skyrmions is maximized in the irradiated device while no skyrmions are formed in the nonirradiated device for the same current density. This enables the localization of current \cite{Kern2022} and field \cite{Juge2021, Hu2022} driven skyrmion nucleation in devices. Surprisingly, this is not the case for the largest two doses studied, which suggest that there exists an optimal {\Ga} dose for the localization, which should depend on the material stack and the parameters of the ion beam.

The number of skyrmions that can be nucleated in a device was found to depend linearly on the ion dose. Contrary to the threshold current, the number of skyrmions keeps increasing for the entire dose range. This observation might be interesting for multistate memory applications \cite{Zhang2018, Li2022}, as well as neuromorphic computation applications \cite{Grollier2020, Pinna2020}, since this means that the average number of skyrmions after nucleation in a device can be effectively controlled using {\Ga} ion irradiation. Furthermore, by utilizing the ability to pattern the ion irradiation it would be possible to control the skyrmion density in devices locally, a hitherto unexplored possibility in skyrmion devices. Finally, we showed that the distribution of nucleation currents becomes broader for devices irradiated with a relatively large ion dose. This could be useful in applications where a more precise control of the skyrmion number is required than can be provided by nonirradiated material systems. Concluding, we have shown that {\Ga} ion irradiation can be used to effectively control the threshold current and efficiency of the current-driven nucleation of magnetic skyrmions.

\begin{acknowledgments}
M.C.H.J. and B.H.M.S. contributed equally. This work is part of the Gravitation programme 'Research Centre for Integrated Nanophotonics', which is financed by the Dutch Research Council (NWO). M.J.M. and J.L. acknowledge support as part of the research programme of the Foundation for Fundamental Research on Matter (FOM), which is a part of NWO. This work was supported by the Eindhoven Hendrik Casimir Institute.
\end{acknowledgments}

\nocite{NoteX}

%

\pagebreak
\widetext
\begin{center}
\textbf{\large Supplemental Materials: Controlling magnetic skyrmion nucleation with Ga$^{+}$ ion irradiation}
\end{center}
\setcounter{section}{0}
\setcounter{equation}{0}
\setcounter{figure}{0}
\setcounter{table}{0}
\setcounter{page}{1}
\makeatletter
\renewcommand{\thesection}{S\Roman{section}}
\renewcommand{\theequation}{S\arabic{equation}}
\renewcommand{\thefigure}{S\arabic{figure}}
\renewcommand{\thetable}{S\arabic{table}}
\renewcommand{\bibnumfmt}[1]{[S#1]}
\renewcommand{\citenumfont}[1]{S#1}

\section{Variation in the threshold current between devices}
\begin{figure*}[ht]
\includegraphics{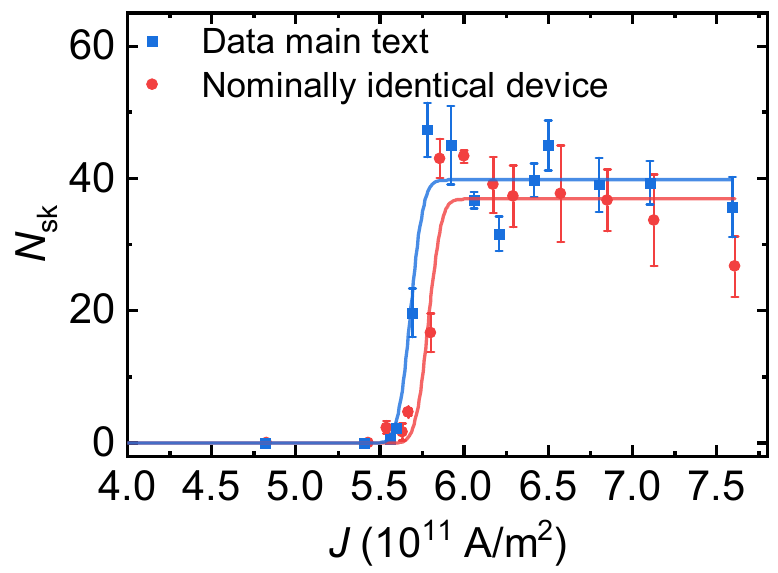}
\caption{\label{fig:Figure_S1} The measurement shown in \cref{fig:Figure_3} in the main text is shown again in blue. The red data points and fit with \cref{eq:error_function} from the main text is data obtained by performing the same measurement of a different, but nominally identical device.}
\end{figure*}

To confirm that the change in the nucleation-threshold current measured in the main text is not caused by naturally occurring differences between devices we have repeated the experiment highlighted in \cref{fig:Figure_3} of the main text on a different but nominally-identical device grown during a second deposition run. Both data sets are plotted in \cref{fig:Figure_S1}. We find that there is a small difference in the nucleation threshold of $2{\%}$ relative to the threshold current reported in the main text. This is an order of magnitude lower than the observed change in the threshold current due to {\Ga} ion irradiation for the highest doses and we therefore attribute the observed changes to the effect of the {\Ga} ions.

\section{Skyrmion nucleation as a function of field and current density}
\begin{figure*}[ht]
\includegraphics[width=\textwidth]{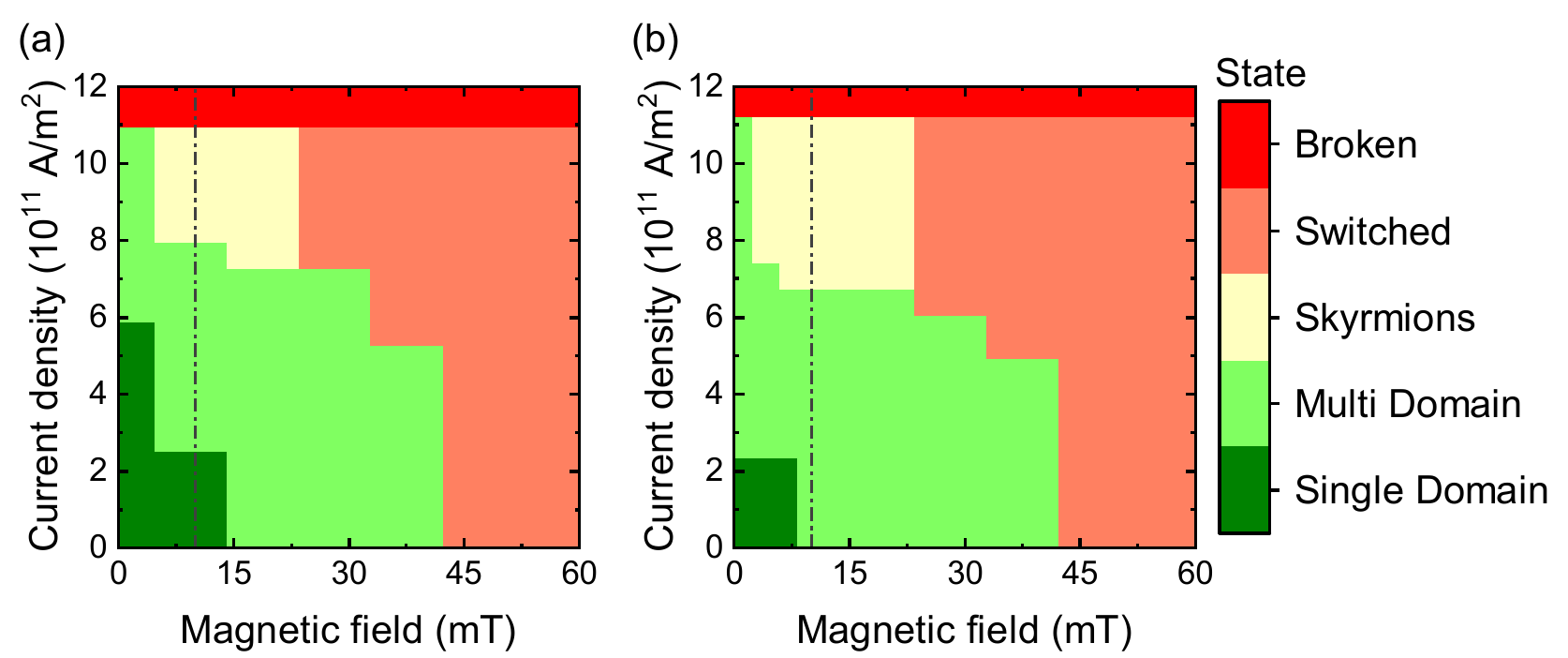}
\caption{\label{fig:Figure_S2} Phase diagrams showing the final magnetization state after applying a sequence of 2000 bipolar current pulses (pulse length: \SI{10}{ns}, delay between pulses: \SI{10}{\micro\second}). Before the pulses are applied the magnetization is saturated in the negative z-direction, after which a bias field is applied in the positive z-direction (the magnetic field on the horizontal axis). This is done for two different devices: (a) a nonirradiated device and (b) a device which has been irradiated with a dose of $d=10\times10^{12}$ {\Ga} ions \SI{}{cm^{-2}}. The dashed black lines indicate the field where the measurements in the main text are performed.}
\end{figure*}

In this section, we show how the nucleation of skyrmions in our devices depends on both the applied-magnetic field and the current density. This measurement is performed with a different pulse generator than in the main text (Picosecond Pulse Labs model A10,070A). The length of the applied current pulses is \SI{10}{ns} and the time between pulses is \SI{10}{\micro\second} to prevent cumulative heating. The results are summarized in \cref{fig:Figure_S2} for a nonirradiated device in (a) and a device irradiated with $d=10\times10^{12}$ ions \SI{}{cm^{-2}} in (b). Plotted is the phase diagram where the final state in the device has been labelled for each pair of current density and applied bias field. We distinguish between five different cases: (i) No change from the initial saturated state (magnetization anti-parallel to the applied field); (ii) a multidomain state without skyrmions; (iii) one or more skyrmions in the device (can be in combination with larger worm-like domains); (iv) a device where the magnetization is aligned with the applied field everywhere; and (v) a broken device. 

For small magnetic-bias fields and low current densities the magnetization of the devices is not affected by the applied current pulses. If either the current density or the field is increased the magnetization becomes sufficiently unstable and a multidomain state is formed. For even larger magnetic fields the magnetization switches fully and aligns parallel to the applied bias field. For a small range of low bias fields and high current densities a state with skyrmions can be reliably reached in both the nonirradiated device and the irradiated device. We note that the phase diagrams shown here are qualitatively similar to those measured in Ref. \cite{Sup_Scarioni2021}.

Although the overall characteristics of the two phasediagrams is similar there are several differences. (i) The magnetization of the irradiated device is less stable at zero field, where the current density required to reach a multidomain state is significantly reduced. (ii) The switched state is reached with lower current densities for fields below the coercive field in the case of the irradiated device. (iii) The minimum current density where skyrmions form in the devices is lower in the irradiated device compared to the nonirradiated device, for all bias fields. All three of these observations indicate that changes in the magnetization are easier in the case of the irradiated device compared to the nonirradiated device. Finally, the nucleation of skyrmions appears to occur at lower fields in \cref{fig:Figure_S2}(b) compared to (a). However, the resolution of the field axis is higher in \cref{fig:Figure_S2}(b) compared to (a) and therefore this cannot be used as proof that skyrmions can be nucleated at smaller bias fields in irradiated devices.

\section{MFM data and counting skyrmions}\label{sec:counting}
All MFM images in the main paper are processed using the Gwyddion software. We first crop the data so that only the narrow current line of the devices remains. Next, a plane is fitted to several regions of the data belonging to the same domain and then subtracted from the data. Next, we correct for small horizontal scars using the Gwyddion function with the same name and finally a Gaussian blur is applied with a 2 px standard deviation. The original size of all MFM images is \SI{10}{\micro\metre} $\times$ \SI{2.5}{\micro\metre} and the number of pixels is $1024 \times 256$ pixels, making the size of each pixel approximately \SI{9.8}{nm} $\times$ \SI{9.8}{nm}.

To count the number of skyrmions in our MFM images we use a procedure based on Ref. \cite{Sup_Tan2020}. Using MatLab we apply a Gaussian blur with a sigma of 3.5 px to the MFM data before binarizing the image using Imbinarize() with an adaptive threshold. Skyrmions in all our data sets appear as white regions in this binarized image. Regions that are smaller than 50 pixels are discarded to remove noise from the binarized images. All the white regions in the binarized image are then collected using the bwboundaries() function and analyzed using regionprops() to find the centroid position, circularity as well as the minor and major axis lengths.

\begin{figure*}[ht]
\includegraphics{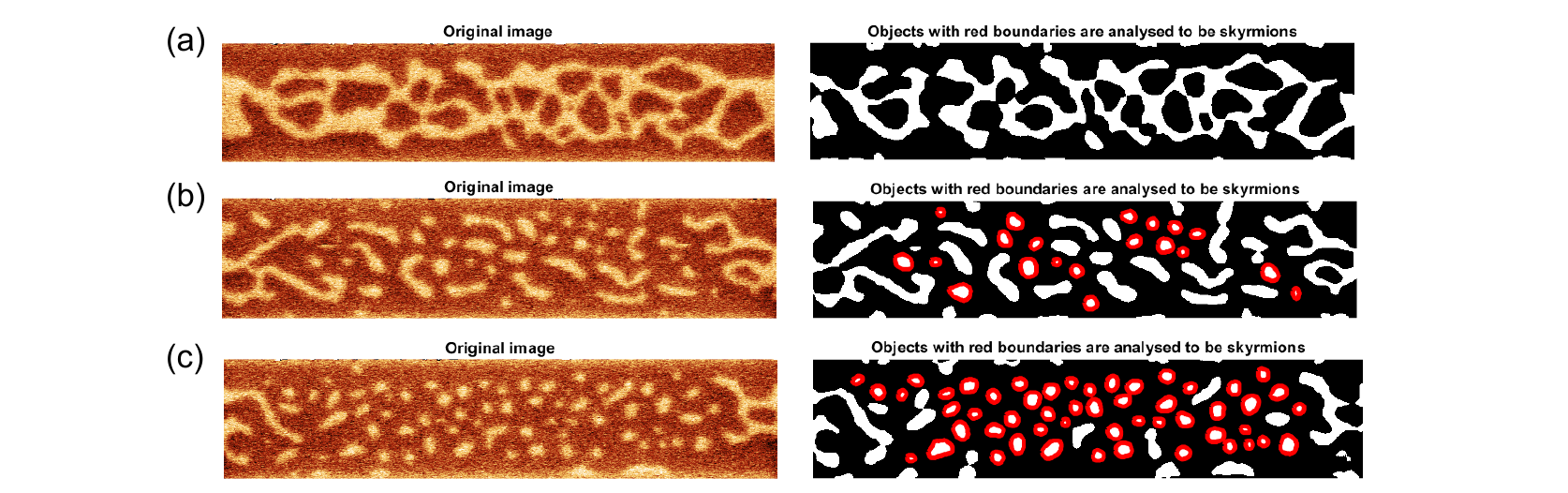}
\caption{\label{fig:Figure_S3} The measurement shown in \cref{fig:Figure_3}(b-d) analyzed using the script outlined above. The regions circulated in red in the black and white figures are counted as skyrmions.}
\end{figure*}

Next, we loop through all the regions and test for the following conditions to determine if it should be considered as a skyrmion:

\begin{itemize}
\item The major axis length $l_{\text{M}}$ must be within the following interval
\begin{equation*}
50 \text{nm} < l_{\text{M}} < 2 * R_{\text{S}},
\end{equation*}
\noindent where $R_{\text{S}}$ is the radius predicted by the model of B{\"u}ttner {\ea} in Ref. \cite{Sup_Buettner2018}. This length was chosen to eliminate objects too large to be considered skyrmions and because we observed that most of the circular domains were smaller than would be predicted by this model. The minimum length is chosen to exclude artifacts of the binarization process.
\item The center of the region (the centroid) must be further that 15 pixels from the edge of the scan. This is used to eliminate regions that are attached to the edges of the device.
\item The circularity of the region $C$, defined as the ratio between $4 \pi A$ ($A$ is the area) and the circumference squared and one for a perfect circle, should satisfy $C > 0.6$.
\item The ratio of the major and minor axis $r$ (the longest and shortest straight line through the region) should satisfy $r < 1.75$
\end{itemize}

The cut-off value for the last two conditions were determined by Tan {\ea} in Ref. \cite{Sup_Tan2020} to maximize the accuracy of the counting procedure. In \cref{fig:Figure_S3} we show several examples of this procedure. The domains that are counted as skyrmions are labelled in red.

\section{MFM images corresponding to Fig. 3 in the main text}
\begin{figure*}
\includegraphics[width=\textwidth]{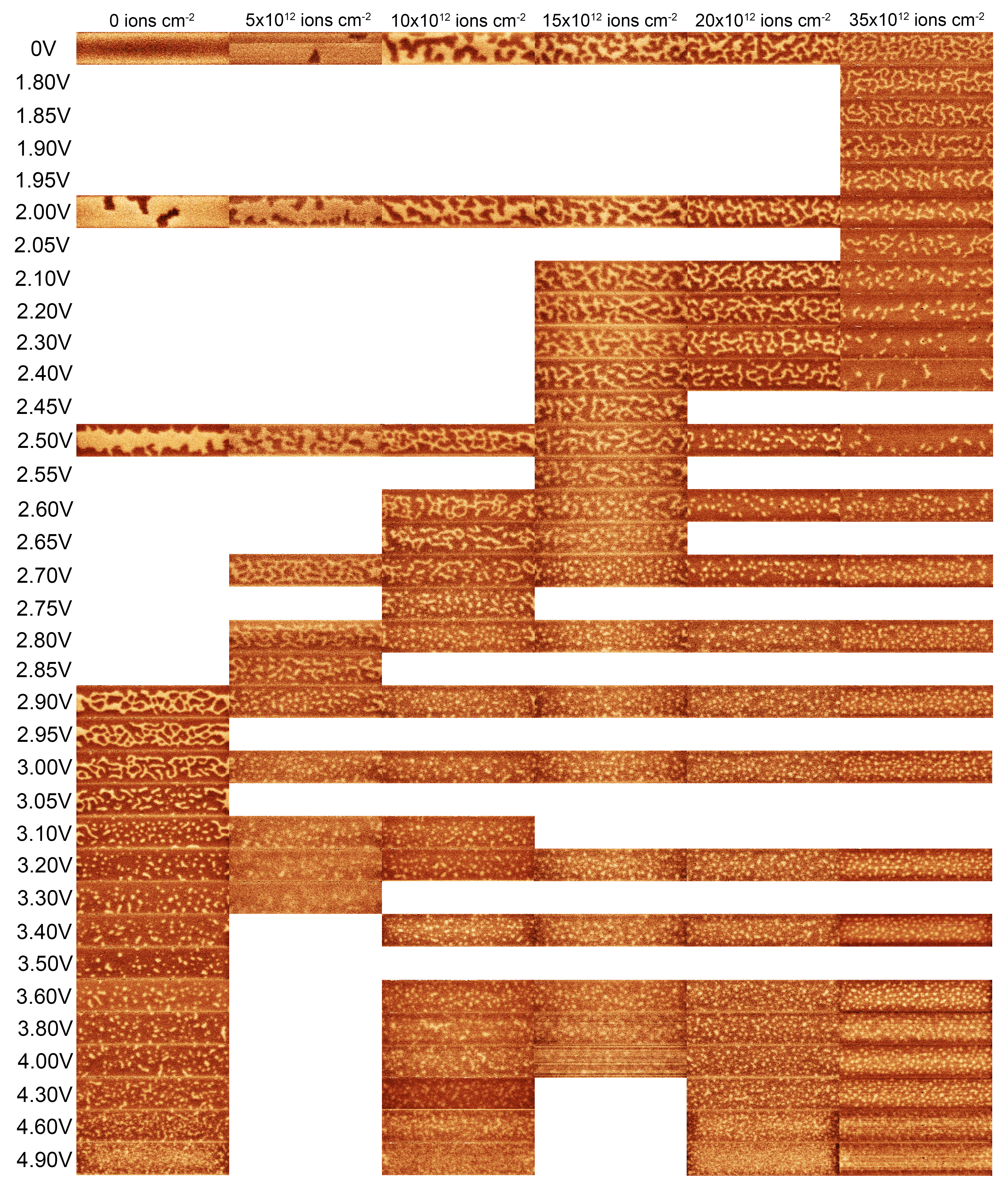}
\caption{\label{fig:Figure_S4} MFM images corresponding to \cref{fig:Figure_4} in the main text.}
\end{figure*}

In \cref{fig:Figure_S4} we show MFM images that correspond to the nucleation events reported in \cref{fig:Figure_4} of the main text. For each combination of dose and current density, we only show one of the three scans that were used in the analysis. The other two scans were qualitatively similar. Here we report the voltage that was set on the pulse generator instead of the current density to make the position of the MFM scans in the figure easier to determine. Due to small differences in the resistance between devices the current densities corresponding to the voltage can differ by a few percent ($\Delta R / R < 4 \%$). These differences have been taken into account in the main text. 

\begin{figure}
\includegraphics[width=\textwidth]{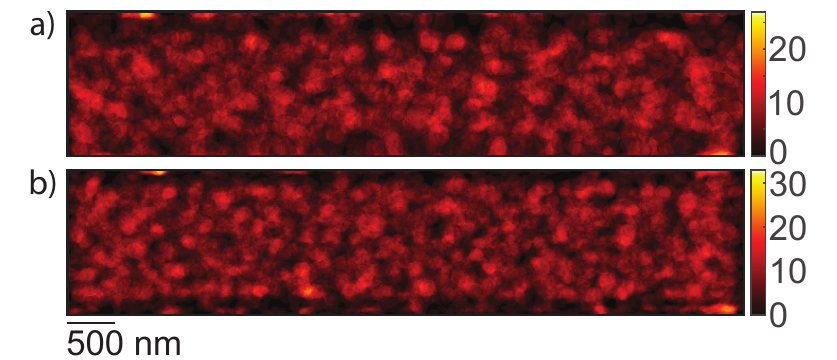}
\caption{\label{fig:Figure_S5}(a) Sum of the binarized MFM images that were used in the analysis of \cref{fig:Figure_4}(a) in the main text, for the nonirradiated device. (b) The same plot but for the device irradiated with $d=35\times10^{12}$ ions \SI{}{cm^{-2}}.}
\end{figure}

In \cref{fig:Figure_S5}(a, b) we also show the sum of the binarized images used in \cref{fig:Figure_4}(a) in the main text for the nonirradiated device and the device irradiated with $d=35\times10^{12}$ ions \SI{}{cm^{-2}}, respectively. Only those images taken for curents larger than the threshold current were used to ensure that the majority of the domains are skyrmions. The colour of each pixel in such an image indicates how often a skyrmion was present on that pixel after nucleation. Apart from some preferential nucleation sites the domains and skyrmions appear to generate randomly throughout the devices.

\newpage

\end{document}